\documentclass[
showpacs,
floatfix,
aps,
prl,
twocolumn,
superscriptaddress,
amssymb,
]{revtex4-1}
\usepackage{times}
\usepackage{amssymb}
\usepackage{latexsym}
\usepackage[dvips]{graphicx}
\usepackage{amsmath}
\usepackage{float}

\usepackage{graphicx}
\usepackage{dcolumn}
\usepackage{amsfonts}
\usepackage{bm}

\usepackage{mathtools}

\newcommand{\be}{\begin{equation}}
\newcommand{\ee}{\end{equation}}
\newcommand{\bea}{\begin{eqnarray}}
\newcommand{\eea}{\end{eqnarray}}

\def\rxx{R_{xx}}
\def\ryy{R_{yy}}
\def\rxy{R_{xy}}

\def\rc{R_{\mbox{\scriptsize {c}}}}

\def\easy{\left < 110 \right >}

\def\easy{\left < 110 \right >}

\def\nus{\nu^\star}

\newcommand{\req}[1]{Eq.\,(\ref{#1})}

\newcommand{\rfig}[1]{Fig.\,\ref{#1}}

\newcommand{\rref}[1]{Ref.\,\onlinecite{#1}}
\newcommand{\rrefs}[2]{Refs.\,\onlinecite{#1},\,\onlinecite{#2}}

\begin{document}
\title{Two- and three-electron bubbles in Al$_{x}$Ga$_{1-x}$As/Al$_{0.24}$Ga$_{0.76}$As quantum wells}

\author{X. Fu}
\affiliation{School of Physics and Astronomy, University of Minnesota, Minneapolis, Minnesota 55455, USA}
\author{Q.~Shi$^\dagger$}
\affiliation{School of Physics and Astronomy, University of Minnesota, Minneapolis, Minnesota 55455, USA}
\author{M.~A.~Zudov}
\email[Corresponding author: ]{zudov001@umn.edu}
\affiliation{School of Physics and Astronomy, University of Minnesota, Minneapolis, Minnesota 55455, USA}
\author{G.\,C. Gardner}
\affiliation{Microsoft Quantum Lab Purdue, Purdue University, West Lafayette, Indiana 47907, USA}
\affiliation{Birck Nanotechnology Center, Purdue University, West Lafayette, Indiana 47907, USA}
\author{J.\,D. Watson$^\ddagger$}
\affiliation{Birck Nanotechnology Center, Purdue University, West Lafayette, Indiana 47907, USA}
\affiliation{Department of Physics and Astronomy, Purdue University, West Lafayette, Indiana 47907, USA}
\author{M.\,J. Manfra}
\affiliation{Microsoft Quantum Lab Purdue, Purdue University, West Lafayette, Indiana 47907, USA}
\affiliation{Birck Nanotechnology Center, Purdue University, West Lafayette, Indiana 47907, USA}
\affiliation{Department of Physics and Astronomy, Purdue University, West Lafayette, Indiana 47907, USA}
\affiliation{School of Electrical and Computer Engineering and School of Materials Engineering, Purdue University, West Lafayette, Indiana 47907, USA}
\received{27 February 2019; published 3 April 2019}

\begin{abstract}
We report on transport signatures of eight distinct bubble phases in the $N=3$ Landau level of a Al$_{x}$Ga$_{1-x}$As/Al$_{0.24}$Ga$_{0.76}$As quantum well with $x = 0.0015$.
These phases occur near partial filling factors $\nus \approx 0.2\,(0.8)$ and $\nus \approx 0.3\,(0.7)$ and have $M = 2$ and $M = 3$ electrons (holes) per bubble, respectively.
We speculate that a small amount of alloy disorder in our sample helps to distinguish  these broken symmetry states in low-temperature transport measurements.

\end{abstract}

\maketitle

While the effect of disorder on transport characteristics of a two-dimensional electron gas (2DEG) is usually deemed detrimental, there exist many situations in which the disorder is beneficial. 
The most celebrated examples are integer \citep{klitzing:1980} and fractional \citep{tsui:1982} quantum Hall effects (QHEs) which rely on single-(quasi)particle localization by the disorder potential.
Many nonequilibrium transport phenomena in very high Landau levels, such as microwave-\citep{zudov:2001a} and Hall field-induced resistance oscillations \citep{yang:2002}, along with several other related phenomena \citep{khodas:2010,shi:2017a}, also benefit from a modest amount of impurities which can provide large-angle scattering.

Furthermore, disorder provides a pinning potential for Wigner crystals \citep{wigner:1934,lozovik:1975,yoshioka:1979,fukuyama:1979,andrei:1988,willett:1988,goldman:1988,li:2010} and ``bubble'' phases \citep{koulakov:1996,fogler:1996,moessner:1996,lilly:1999a,du:1999,cooper:1999,eisenstein:2002,chen:2019} allowing for their transport manifestation. 
These bubble phases can be viewed as generalizations of a Wigner crystal formed from clusters of $M \ge 1$ particles per unit cell. 
Such clustering of electrons (or holes) into ``bubbles'' is made possible in partially-filled high Landau levels because ring-like electron wavefunctions interact with a box-like potential which is a result of an interplay between long-range direct and short-range exchange components of Coulomb interaction \citep{koulakov:1996}.
At low temperatures these $M$-particle bubbles crystallize into a triangular lattice with a lattice constant $\Lambda_b \approx 3.3 \rc$ \citep{fogler:1996}, where $\rc = l_B\sqrt{2N+1}$ is the cyclotron radius, $N$ is the Landau level index, $l_B = (\hbar/e B)^{1/2}$ is the magnetic length, and $B$ is the perpendicular magnetic field.
Being pinned by disorder, such bubble crystals are insulating and the measured resistances are akin to those at the nearest integer filling factors $[\nu]$, i.e., both $\rxx$ and $\rxy$ are small, while $\rxy$ exhibits integer QHE.
This picture is also supported by the observation of pinning mode resonances in microwave spectroscopy studies \citep{lewis:2002,lewis:2004}.

To date, experiments on the bubble phases have focused primarily on $N = 1$ \citep{eisenstein:2002,xia:2004,csathy:2005,deng:2012a,baer:2015,rossokhaty:2016,shingla:2018,bennaceur:2018} and $N = 2$ \citep{lilly:1999a,du:1999,cooper:1999,gores:2007,deng:2012b,wang:2015,friess:2018,bennaceur:2018,chen:2019} Landau levels.
At $N = 1$, experiments revealed signatures of eight bubble phases occurring at $\nus \approx 0.29$ and $\nus \approx 0.43$ (see, e.g., \rref{deng:2012a}) in each spin sublevel (as well as their electron-hole symmetric values, $\nus \approx 1-0.29$ and $\nus \approx 1-0.43$), where $\nus = \nu - \lfloor \nu \rfloor$ is  the partial filling of the Landau level and $\lfloor \nu \rfloor = \max \{ m \in  \mathbb{Z}\,|\,m \le \nu$\} is the integral part of $\nu$.
These states can be ascribed to one- and two-particle bubbles, respectively \citep{goerbig:2003,goerbig:2004}.
At $N = 2$, transport studies (see, e.g., \rref{deng:2012b}) found four insulating states accompanied by integer QHE near $\nus \approx 0.28$ and $\nus \approx 1-0.28$, which likely reflect formation of bubble crystals with $M = 2$ \citep{note:5}.
While at $N = 2$ theory (see, e.g., \rrefs{shibata:2001}{goerbig:2004}) also predicts bubble phases  with $M = 1$, to our knowledge, their existence has not been confirmed in transport measurements \citep{note:2}.
Similar to $N = 2$, theory \citep{goerbig:2004,yoshioka:2002} predicts at least two kinds of bubbles at $N = 3$, with $M = 2$ and $M = 3$, but experiments have so far detected only four isotropic insulating states centered around $\nus \approx 0.27$ (see, e.g., \rref{shi:2016b}).

In this Rapid Communication we report on transport signatures of \emph{eight} distinct bubble phases in the $N = 3$ Landau level of a Al$_{x}$Ga$_{1-x}$As/Al$_{0.24}$Ga$_{0.76}$As quantum well with $x = 0.0015$.
These signatures are observed in both lower and upper spin branches near partial filling factors $\nus \approx 0.2$ and $\nus \approx 0.3$ (and their particle-hole conjugates $\nus \approx 0.8$ and $\nus \approx 0.7$), which correspond to $M = 2$ and $M = 3$ electrons (or holes) per bubble, respectively.
The temperature dependence suggests that three-particle bubbles start to develop at somewhat higher temperature than two-particle bubbles.
The data in the control sample (with $x = 0$) on the other hand, show only four insulating states which, however, extend over wider ranges of $\nus$, i.e. $0.20 \lesssim \nus \lesssim 0.33$.
We believe that a small amount of alloy disorder helps to distinguish between two- and three-particle bubbles in our Al$_{x}$Ga$_{1-x}$As/Al$_{0.24}$Ga$_{0.76}$As quantum well.

While we have observed signatures of two- and three-particle bubbles in several 30 nm-wide Al$_{x}$Ga$_{1-x}$As/Al$_{0.24}$Ga$_{0.76}$As quantum wells (with identical heterostructure design but with different Al content $x$ from 0.0 to 0.0036 \citep{gardner:2013}), here we present the data obtained from a sample with $x = 0.0015$. 
After a brief low-temperature illumination, our sample had the density $n_e \approx 2.9 \times 10^{11}$ cm$^{-2}$ and the mobility $\mu \approx 3.6 \times 10^6$ cm$^2$V$^{-1}$s$^{-1}$.
The sample was a $4 \times 4$ mm square with eight indium contacts positioned at the corners and the midsides. 
Resistances, $\rxx$, $\ryy$, and $\rxy$ were measured using a four-terminal, low-frequency lock-in technique.

\begin{figure}[t]
\centering
\includegraphics{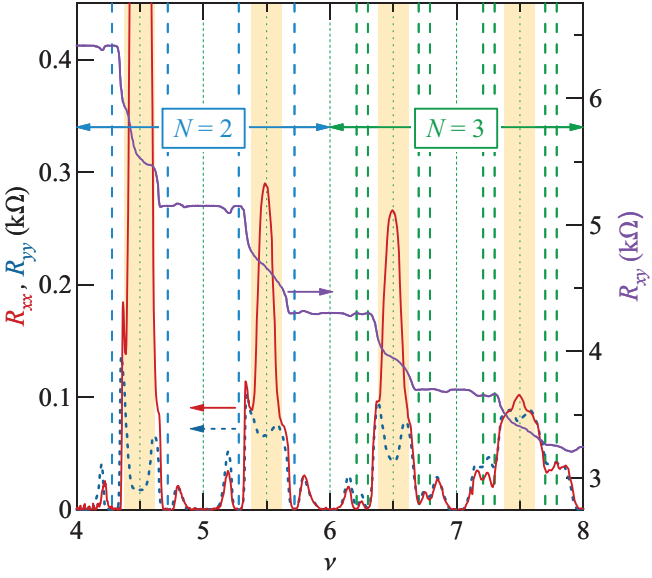}
\vspace{-0.15 in}
\caption{(Color online)
Longitudinal resistance $\rxx$ (solid line, left axis), $\ryy$ (dotted line, left axis), and Hall resistance $\rxy$ (right axis) as a function of the filling factor $\nu$ at $T \approx 25$ mK. 
Bubble phases in the $N=2$ and $N=3$ Landau levels are marked by vertical dashed lines drawn at $\nus = 0.28,  0.72$ and at $\nus = 0.21, 0.30, 0.70, 0.79$, respectively.
Shaded regions correspond to $0.38 \le \nus \le 0.62$, where stripe phases form (see, e.g., \rref{shibata:2001}).
}
\vspace{-0.15 in}
\label{fig1}
\end{figure}
In \rfig{fig1} we present the longitudinal resistances $\rxx$ (solid line, left axis), $\ryy$ (dotted line, left axis), and the Hall resistance $\rxy$ (right axis) as a function of the filling factor $\nu$  measured at $T \approx 25$ mK. 
The shaded areas mark the regions $0.38 \le \nus \le 0.62$ where $\rxx > \ryy$ reflecting the formation of anisotropic stripe phases \citep{koulakov:1996,lilly:1999a,du:1999} with the easy direction along the $\easy$ crystal axis.
In the $N = 2$ Landau level the data clearly show four isotropic insulating states occurring near partial fillings $\nus \approx 0.28$ and $\nus \approx 0.72$ (marked by vertical dashed lines) of both the lower and the upper spin branch. 
These states are attributed to the formation of bubble crystals formed by clusters of $M=2$ electrons or holes. 
As expected, $\rxx  \approx  \ryy \approx 0$, while $\rxy$ exhibits re-entrant QHE at $\rxy = R_{\rm K}/[\nu]$, where $R_{\rm K} = h/e^2 \approx 25.812$ k$\Omega$ is the von Klitzing constant.

\begin{figure}[t]
\centering
\includegraphics{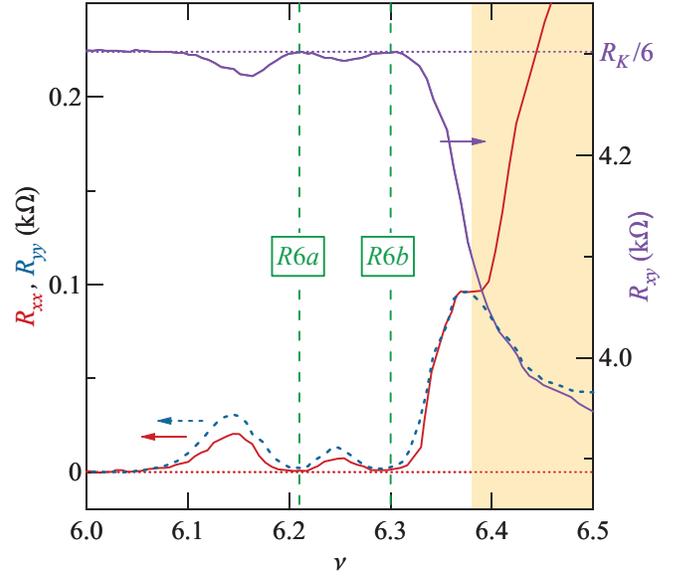}
\vspace{-0.15 in}
\caption{(Color online)
Zoom-in view of \rfig{fig1} for $\nu$ between 6.0 and 6.5. 
Re-entrant integer quantum Hall states are marked by $R6a$ and $R6b$.
}
\vspace{-0.15 in}
\label{fig2}
\end{figure}

Remarkably, $\rxx$ and $\ryy$ in the $N = 3$ Landau level reveal \emph{eight} well-defined minima, two on each side of both half-filled spin sublevels.
The positions of these minima are marked by vertical dashed lines drawn at $\nus = 0.21, 0.30, 0.70, 0.79$.
Since two of these partial fillings are fairly close to $\nus = 1/5$ and $\nus = 4/5$ which, in principle \citep{note:3}, might support QHE, it is important to examine the $\rxy$ data more closely.
In \rfig{fig2} we present a zoom-in view of the data for $6.0 < \nu < 6.5$. 
One observes that as both $\rxx$ and $\ryy$ approach zero at $\nus \approx 0.21$ and $\nus \approx 0.30$, the Hall resistance $\rxy$ exhibits re-entrant integer QHE with $\rxy = R_{\rm K}/6$ and not fractional QHE.
These observations strongly suggest the formation of bubble phases at these filling factors, which we label $R6a$ and $R6b$.

Although the remaining six minima do not reach zero in our experiment, they (i) occur either near the same partial fillings $\nus$ or their electron-hole symmetric counterparts $\nus = 0.79$ and $\nus = 0.70$, and (ii) are accompanied by re-entrant QHE features in the $\rxy$.
We thus believe that these features also signal formation of the bubble phases and we will refer to them as $R6c,R6d,R7a,R7b,R7c$, and $R7d$.
As illustrated in \rfig{fig1}, partial fillings of bubble phases show little difference between the lower and the upper spin branches (i.e., $\nus_{R6\alpha} \approx \nus_{R7\alpha}$ for $\alpha = a,b,c,d$), and, as already mentioned, are electron-hole symmetric (i.e., $\nus_{Rid} \approx 1 - \nus_{Ria}$ and $\nus_{Ric} \approx 1 - \nus_{Rib}$ for $i = 6,7$).

We can estimate the number $M$ of electrons per bubble from $\nus < 1/2$, Landau level index $N$, and the lattice constant of the bubble phase $\Lambda_b$ using \citep{note:eq}
\be
M = \frac {\sqrt{3}}{2\pi} \left ( \frac {\Lambda_b} {\rc}\right )^2 (N+1/2) \nus\,.
\label{eq.m}
\ee
With $\Lambda_b \approx 3.3 \rc$ \citep{fogler:1996,fogler:1997}, $N=3$, and $\nus = 0.21,0.30$, we find $M \approx 2$ for $Ria$ and $M \approx 3$ for $Rib$ \citep{note:1}.
These values are in excellent agreement with the theory \citep{fogler:1997,yoshioka:2002,goerbig:2004} predicting formation of bubble phases with $M = 2$ and $M = 3$ electrons per bubble in the $N = 3$ Landau level.
We thus conclude that $Ria,Rid$ and $Rib,Ric$ are two- and three-particle bubble phases, respectively.

\begin{figure}[t]
\centering
\includegraphics{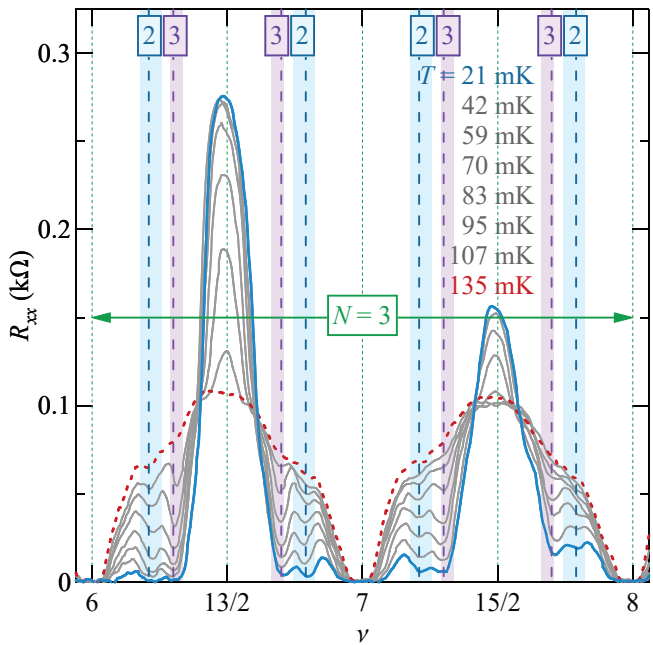}
\vspace{-0.15 in}
\caption{(Color online)
Longitudinal resistance $\rxx$ vs filling factor $\nu$ in the $N = 3$ Landau level at different temperatures from 21 mK (thick solid line) to 135 mK (dotted line), as marked.
Shaded areas correspond to the ranges of $\nus$ where calculations \citep{yoshioka:2002} predict bubble phases with $M=2$ and $M=3$, as marked.
}
\vspace{-0.15 in}
\label{fig3}
\end{figure}

To further test the idea that our data manifest the formation of the bubble phases, we have examined the temperature dependence.
In \rfig{fig3}(a) we present longitudinal resistance $\rxx$ as a function of the filling factor $\nu$ measured at different temperatures $T$ from 21 mK (thick solid line) to 135 mK (dotted line), as marked.
At the highest $T \approx 135$ mK, the $\rxx$ is rather featureless, apart from QHEs near integer $\nu$.
In the vicinity of $\nus = 1/2$, the $\rxx$ rapidly rises with decreasing $T$ reflecting formation of stripe phases.
Away from half-filling, however, the $\rxx$ drops as the temperature is lowered and double minima develop on each side of the half-filling.
These minima remain roughly at the same filling factors (marked by vertical dashed lines) over the entire temperature range.
Moreover, these filling factors fall within the ranges of $\nus$ (shaded areas) where density matrix renormalization group calculations \citep{yoshioka:2002} predict bubble phases with $M=2$ and $M=3$. 

Further examination of the data in \rfig{fig3} shows that the minima near $\nus \approx 0.3 (0.7)$ develop faster with decreasing $T$ than the ones near $\nus \approx 0.2 (0.8)$, a behavior most evident at intermediate temperatures, although eventually both approach roughly the same resistance values at the lowest $T$.
Understanding this subtle difference in the temperature dependencies of the two phases will require further investigations.

\begin{figure}[t]
\centering
\includegraphics{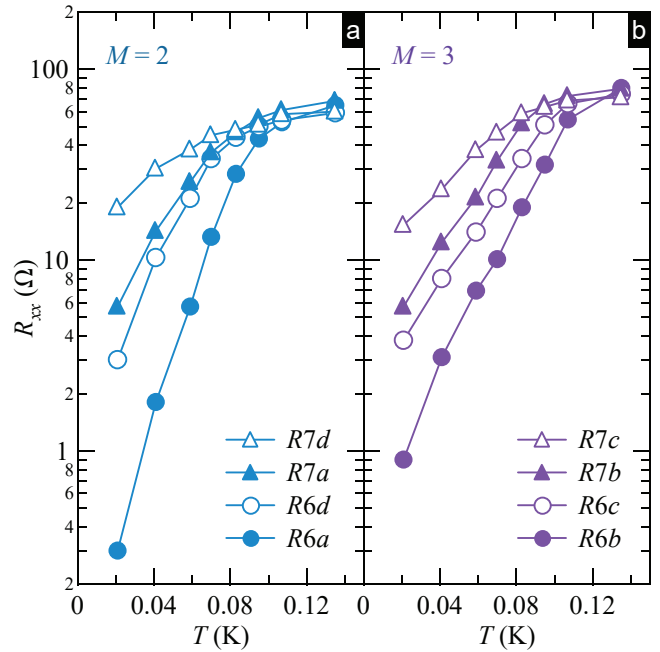}
\caption{(Color online)
Resistance $\rxx$ at $\nus$ corresponding to bubble phases in the $N = 3$ Landau level with (a) $M = 2$ and (b) $M=3$ particles per bubble, see legend, as a function of temperature $T$. 
Log-linear scale is used for clarity.
}
\vspace{-0.15 in}
\label{fig4}
\end{figure}

In \rfig{fig4} we plot longitudinal resistance $\rxx$ at filling factors $\nu$ corresponding to bubble phases with (a) $M = 2$ and (b) $M=3$ (as noted in the legend) versus temperature $T$ using the log-linear scale (for clarity).
Both data sets manifest very similar behavior, apart from the above-mentioned better development of the three-particle bubbles at intermediate $T$. 
Each of the data sets shows that the low-temperature resistance at the bubble minima grows with the total filling factor $\nu$, suggesting weakening of these phases with increasing $\nu$.
This observation is qualitatively consistent with the monotonic decrease of the onset temperature of the bubble phases in the $N = 2$ Landau level \citep{deng:2012b,note:4}. 

It is interesting to note that the resistance minima which we associate with two- and three-particle bubble phases are separated by a resistance maximum, suggesting particle delocalization at these $\nus$.
This finding seems to agree with calculations \citep{yoshioka:2002} which did not find bubble phases for $0.25 < \nus < 0.30$ at $N = 3$. 
However, our measurements in the control sample (with $x = 0$) show only four insulating states (see, also \rref{shi:2016b}) which extend over much wider ranges of $\nus$, i.e., $0.20 \lesssim \nus \lesssim 0.33$, at low temperatures \citep{note:6}.
Our observation of finite conductivity near $\nus \approx 0.25$ suggests that alloy disorder narrows the ranges of filling factors where bubble phases with $M=2$ and $M=3$ are insulating, allowing one to resolve them separately. 
If at $\nus \approx 0.25$ the bubble phases with $M = 2$ and $M = 3$ are energetically degenerate, one can expect coexistence of both types of bubbles.
As one crosses this filling factor, electrons (or holes) must hop between different types of bubbles as the new bubble lattice is being formed.
Being short-range, alloy disorder can facilitate such hopping via large-angle scattering events (accompanied by large momentum transfer) leading to finite conductivity near the transition.

In summary, we have observed transport signatures of eight bubble phases in the $N = 3$ Landau level of an Al$_{x}$Ga$_{1-x}$As/Al$_{0.24}$Ga$_{0.76}$As quantum well with $x = 0.0015$.
Analysis shows that these phases, found near partial fillings $\nus \approx 0.2$ and $0.8$ ($\nus \approx 0.3$ and $0.7$) of each spin sublevel, contain $M = 2$ and $M = 3$ electrons (holes) per bubble, respectively.
We speculate that a small amount of alloy disorder in our quantum well allows to distinguish these phases, which tend to merge with each other in samples without alloy disorder.

\begin{acknowledgements}
We thank B. Shklovskii for discussions and G. Jones, S. Hannas, T. Murphy, J. Park, A. Suslov, and A. Bangura for technical support.
The work at Minnesota (Purdue) was supported by the U.S. Department of Energy, Office of Science, Basic Energy Sciences, under Award No. ER 46640-SC0002567 (DE-SC0006671).
A portion of this work was performed at the National High Magnetic Field Laboratory, which is supported by National Science Foundation Cooperative Agreements No. DMR-115749 and No. DMR-1644779 and the State of Florida.
\end{acknowledgements}

\small{$^\dagger$Department of Physics, Columbia University, New York, NY, 10027;}
\small{$^\ddagger$Present address: Microsoft Station-Q at Delft University of Technology, 2600 GA Delft, The Netherlands.}


\begin{thebibliography}{49}
\bibitem[{not({\natexlab{g}})}]{note:6}
\bibinfo{note}{As the temperature is raised, this range of $\nus$ shrinks
  asymmetrically towards $\nus \approx 0.3$, a point where we observe $M = 3$
  bubble phase in samples with non-zero $x$.}
\end{thebibliography}

\end{document}